\documentclass[a4paper,fleqn,usenatbib]{mnras}

\usepackage{newtxtext,newtxmath}

\usepackage[T1]{fontenc}
\usepackage{ae,aecompl}

\usepackage{graphicx}	
\usepackage{amsmath}	
\usepackage{amssymb}	
\usepackage{epstopdf} 
\usepackage{gensymb}
\usepackage{siunitx}
\usepackage[dvipsnames]{xcolor}

\title[Photometry and Spectroscopy of massive stars]{Photometry and spectroscopy of massive stars observed during K2 Campaign 8}

\author[J. M. Eidam et al.]{
J\'essica M. Eidam,$^{1,2}$\thanks{E-mail: jessyeidam@gmail.com}
Laerte Andrade,$^{1}$
Marcelo Emilio,$^{1,2,3}$\newauthor
~M. Cristina Rabello-Soares,$^{4}$
Alan W. Pereira,$^{1,2}$
Eduardo Janot-Pacheco$^{5}$ and\newauthor
~James D. Armstrong$^{3}$
\\
$^{1}$Universidade Estadual de Ponta Grossa, 84030-900 Ponta Grossa, PR, Brazil\\
$^{2}$Observat\'orio Nacional, MCTIC, 20921-400 Rio de Janeiro, RJ, Brazil\\
$^{3}$Institute for Astronomy, University of Hawaii, Maui, HI 96768, USA\\
$^{4}$Physics Department, Universidade Federal de Minas Gerais, 31270-901 Belo Horizonte, MG, Brazil\\
$^{5}$Instituto de Astronomia, Geof\'isica e Ci\^encias Atmosf\'ericas, Universidade de S\~ao Paulo, 05509-090 S\~ao Paulo, SP, Brazil\\
}

\date{Accepted 2019 August 30. Received 2019 August 29; in original form 2018 December 19}

\pubyear{2019}
\begin{document}
\label{firstpage}
\pagerange{\pageref{firstpage}--\pageref{lastpage}}
\maketitle

\begin{abstract}
We report in this paper spectroscopic and photometric analysis of eight massive stars observed during Campaign 8 of the {\it{Kepler}}/K2 mission from January to March 2016. Spectroscopic data were obtained on these stars at OPD/LNA, Brazil, and their stellar parameters determined using SME. Periodic analyses of the light curves were performed through CLEANEST and PERIOD04 algorithms. Mass, luminosity, and radius of our stars were estimated employing CESAM+POSC grids. Three of our stars show significant periodicity. K2 ID 220679442 and  K2 ID 220532854 have periods linked to the stellar rotation. K2 ID 220532854 has prominent silicon lines (\ion{Si}{ii} 4128--4131 \AA), a characteristic presented in the peculiar class of Ap magnetic main sequence stars. However, in our spectral analysis, this object was found to be an evolved, luminous giant star. K2 ID 220466722 was revealed to be a $\delta$ Scuti variable, and 40 individual frequencies were determined for this star. No significant periodicity was found in the light curves for the remaining stars analyzed in this work, besides the instrumental one.
\end{abstract}




\section{Introduction}

The {\it{Kepler}} spacecraft was launched in March 2009. Its primary mission was to monitor more than 150\,000 stars looking for transiting terrestrial exoplanets (Earth-sized planets in and near the habitable zone of Sun-like stars), detection in the Cygnus-Lyra region. Details are described by \citet{Koch2010} and \citet{Borucki2010}. After the loss of two reaction wheels on board the spacecraft in May 2013, the mission was reconfigured to observe along the ecliptic plane, and renamed K2. Since then, it allows the scientific community to propose targets to observe. The K2 mission entails a series of sequential observing ``Campaigns'' of fields distributed in the ecliptic plane. Sun angle constraints, therefore, limit each campaign to a duration of approximately 80 days.

The {\it{Kepler}} mission was designed primarily for the
study of MS stars in the quest for an Earth-like planet.
However, {\it{Kepler}} data has the potential for study upper main sequence stars
\citep[see, for example,][]{murphy2014}.
In fact, only 0.75\% of the stars observed by its primary mission have
effective temperature larger than ~8300K \citep[][Table 4]{mathur2017}.
Rapidly evolving massive stars are rare. Besides their short lifetimes, the number of stars that form per mass interval per unit area in the Milky Way's disk -- the so-called initial mass function -- is strongly mass-dependent. Moreover, the star density for 2, 4 and 8 solar-mass stars correspond to 
approximately 30\%, 10\% and 3\% of the number of stars with one solar mass, respectively \citep{Carroll06}.
Here we perform a full analysis (photometric and spectroscopic) for stars with effective temperature larger than ~8300 K observed during Campaign 8 of K2 mission. These stars are part of the program proposed by our group, where our choice of targets were all stars in the SIMBAD \citep{wenger2000} database that are in the field of view (FoV) of Campaign 8, have spectral types A or B, and magnitude V < 16.
The last criterion was necessary to obtain their ground-based spectra as planned.
Campaign 8 was chosen for our analysis because its FoV is $50\degree$ to $59\degree$ south of the Galactic Plane (and galactic longitude around $130\degree$), whereas the primary mission FoV is centered on galactic coordinates ($76.5\degree, +13.3\degree$). In spite of being a region of lower star density, the C8 FoV has been less studied.

Campaign 8 was observed between January and March 2016, for a total of 78.73 days.
Among the targets there are eight A/B9-type stars (Table \ref{tab:target_list}) proposed by our research group and observed with a 30-minute cadence.
On February 1st, the spacecraft lost fine point control for approximately 30 hours. Coarse point data during this period was removed from our light curves, introducing gaps. Every 6 hours the drift motion of the spacecraft was corrected. This movement causes a systematic signal that is present in the K2 simple aperture photometric light curves. This signal corresponds to the frequency $f$\,=\,4 d$^{-1}$.

A-type stars show a wide range of specificities. More than 30\% exhibit chemical peculiarities \citep{Gray2009book}, making them interesting objects. The most common peculiarities are the Am and Ap phenomena \citep{murphy2014}. In this work, spectroscopy data reveal that two stars, [1] and [3], feature prominent silicon lines (\ion{Si}{ii} 4128-4131 \AA), a characteristic presented by peculiar Ap and Bp stars. Dwarf A-type stars are relatively fast rotators, showing  $v\sin{\textrm{i}}$ significantly over $100$ km s$^{-1}$ \citep{1999A&AS..137..273G}. Slow rotators are also observed among the early-A star. They are the so-called peculiar metallic-line Am stars \citep{Conti1965}.

Variable stars are also found among the A spectral class: $\delta$ Scuti, named after their prototype, are main sequence and giant stars located in the so-called variable instability strip of the HR diagram and are used as standard candles to measure cosmic distances. They are within the mass range of 1.5--2.5 M$_{\sun}$ and spectral type A2 to F5. Most $\delta$ Scuti stars are short-period variables and show pressure (p) nonradial pulsation modes. Some of them are referred to as hybrid stars because they pulsate simultaneously in both $p$ and $g$ (gravity) modes \citep{bowman_kurtz2018}. P-modes are excited by the $\kappa$ mechanism and g-modes by the convective flux blocking mechanism. 

In section~\ref{photometry}, we describe the temporal analysis of the photometric data using the CLEANEST and PERIOD04 algorithms. In section~\ref{spectroscopy}, the spectroscopy analysis using Spectroscopy Made Easy (SME) package is presented. Fundamental parameters such as mass, radius, and luminosity were obtained through the CESAM+POSC grids, and the position of the stars with respect to evolutionary trajectories in the HR diagram could be established. The results of our analyses are presented for each star in section~\ref{results} and a summary of our results in section~\ref{summary}.

\section{Photometric Analysis} \label{photometry}
The {\it{Kepler}} mission provides high photometric accuracy and high temporal resolution. Photometric measurements were taken by the {\it{Kepler}}/K2 mission from January 04 to March 23, 2016. We made use of the K2 Systematic Correction described by \citet{aigran2016} for all the targets in this paper.  Frequency analyses of the eight light curves were made using the CLEANEST \citep{Foster1995} and PERIOD04 \citep{Lenz2005} algorithms.

CLEANEST is an effective analysis tool for detecting signals in time series with irregular time spacing. It uses statistics based on the modified periodogram Lomb-Scargle \citep{1982ApJ...263..835S} and the discrete ``data-compensated'' Fourier transform \citep{1981AJ.....86..619F}. CLEANEST designs the data in an orthogonal subspace of test functions (sine, cosine, and a constant function). The CLEANEST spectrum follows a $\chi^2$ distribution with $r-1$ degrees of freedom, where $r$ is the number of test functions \citep{1996AJ....111..541F}.

PERIOD04 is a Java/C++ tool used in Astronomy to search for frequencies in time series containing gaps. It also allows for estimation of uncertainties in adjusted parameters by employing Monte Carlo simulations.

In order to check if any of the frequencies provided by CLEANEST for each star were related to rotation or variability, we calculated the rotation frequency ($f_\textrm{rot}$) by means of equation~\ref{eq:veq}. For this, v$\sin{i}$ values were obtained for each star by means of the Fourier transform of the magnesium line profile (\ion{Mg}{ii} 4481 \AA) on the stellar spectra (Table~\ref{tab:SME_Media_Robusta_vsini_cesam}). The radii were estimated from spectroscopic data using CESAM $+$ POSC codes (see section~\ref{spectroscopy}). 

\begin{equation} \label{eq:veq}
    f_{\textrm{rot}} = \frac{v_{\textrm{eq}}}{2 \pi R_{\textrm{eq}}}
\end{equation}

\begin{table*}
	\centering
	\caption{Target list. The numbers at the first column are the identification used in figure~\ref{fig:allphotometry} and elsewhere in this paper.}
	\label{tab:target_list}
	\begin{tabular}{lcccccc} 
		\hline
		Number & SIMBAD & K2 ID & RA (J2000) & DEC (J2000) & V mag & V mag Err\\
		\hline
		$[1]$ & BD+10\,102 & 220679442 & 00 52 42.017 & +10 54 50.17 & $9.90$ & $0.03$\\
        $[2]$ & HD\,8919 & 220466722 & 01 28 06.013 & +06 01 11.27 & $10.07$ & $0.04$\\
        $[3]$ & HD\,6164 & 220532854 & 01 02 49.369 & +07 30 02.71 & $7.82$ & $0.01$ \\
		$[4]$ & HD\,9226 & 220280695 & 01 30 54.446 & +02 13 16.53 & $9.51$ & $0.03$\\
        $[5]$ & HD\,3928 & 220607132 & 00 41 55.979 & +09 10 57.71 & $9.48$ & $0.02$\\
        $[6]$ & HD\,6815 & 220631213 & 01 08 55.860 & +09 43 49.82 & $7.29$ & $0.01$ \\
        $[7]$ & HD\,7353 & 220682692 & 01 13 49.570 & +10 59 59.64 & $9.40$ & $0.02$ \\
        $[8]$ & BD+11\,132B & 220731321 & 01 01 02.280 & +12 25 12.97 & $9.93$ & $0.05$ \\
		\hline
	\end{tabular}
\end{table*}

        \begin{figure*}
        	\includegraphics[angle=0.0, trim={1cm 3.63cm 2cm 2.5cm},clip, width=\textwidth]{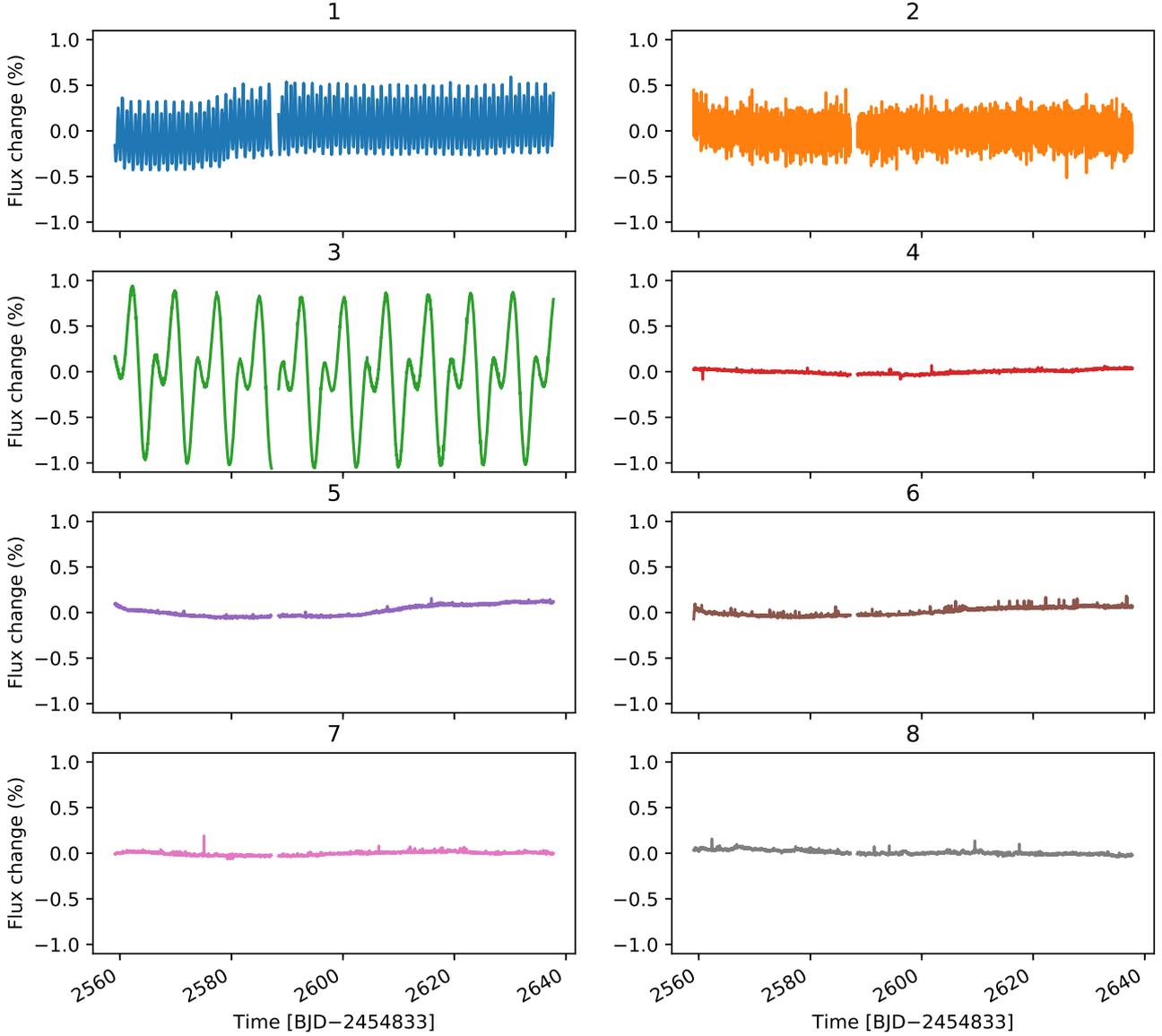}
            \caption{Light curves of the eight stars in this paper treated with the K2SC \citep{aigran2016}. The first three stars show significant variability. The variability of $[1]$ and $[3]$ are compatible with rotation. We identified star $[2]$ as a $\delta$ Scuti variable. The gap around day 2590 is due to the spacecraft loss of fine point on February 1st 2016.}
            \label{fig:allphotometry}
        \end{figure*}

        \begin{figure*}
        	\includegraphics[angle=-90.0,width=\textwidth]{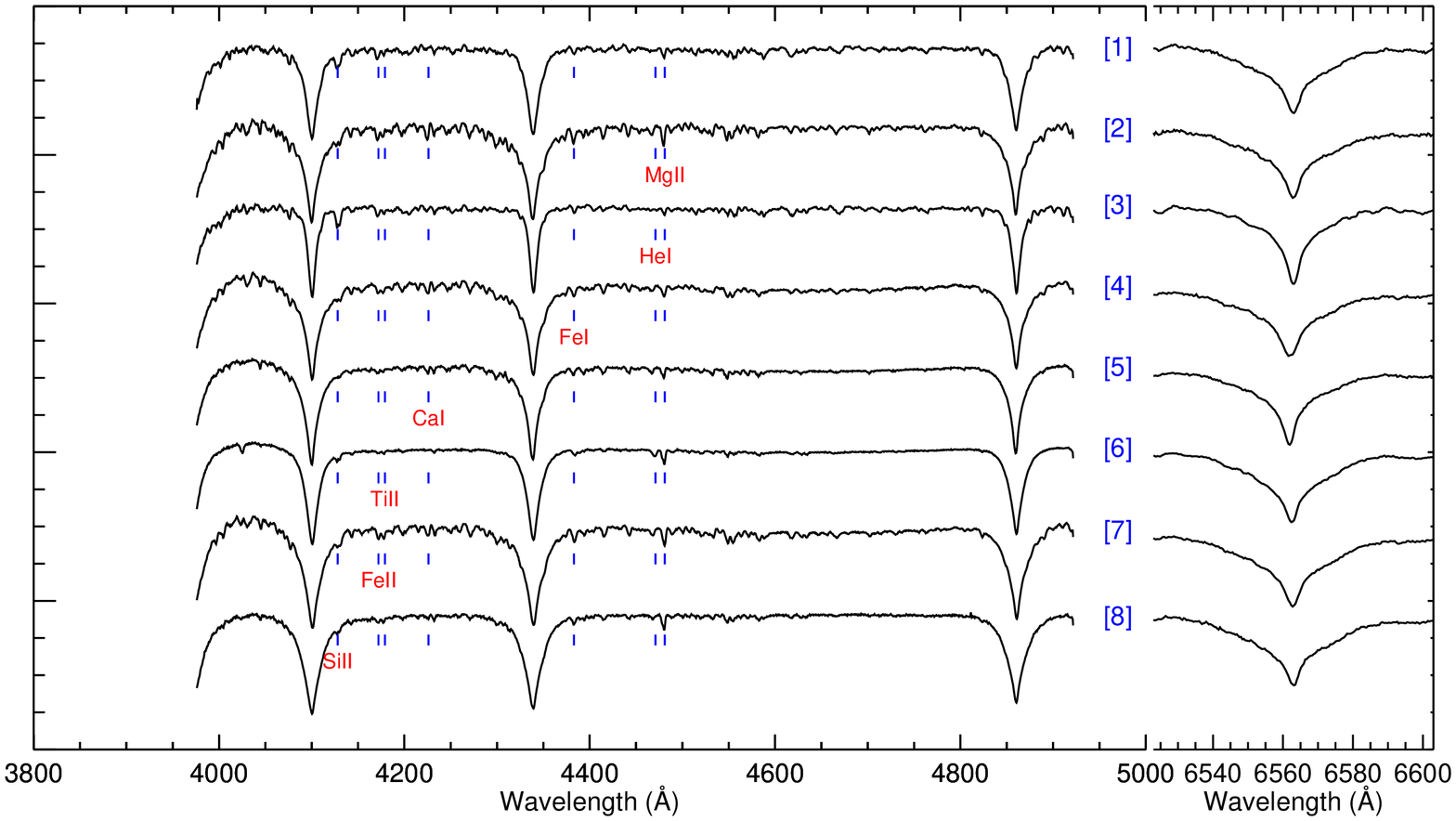}
            \caption{Spectra of the targets in table~\ref{tab:target_list} observed with the 1.6m P-E telescope in OPD/LNA, featuring the spectral classification region (4000--5000 \AA, main lines used identified) and the H$\alpha$ line.}
            \label{fig:spectrablue}
        \end{figure*}

\subsection{Luminosity Distance}

We compared the luminosity found in our spectral analysis with the luminosity derived from Gaia DR2 parallax \citep{gaia2016, gaiaddr2_2018}. The procedure was done following the documentation release 1.1 for Gaia Release 2 (available at \url{http://gea.esac.esa.int/archive/documentation/GDR2/}). We used a geometric distance estimation using first a short distance that varies smoothly as a function of Galactic longitude and latitude according to a Galaxy model made by \cite{bailer2018}. Bolometric corrections were performed using the tables provided by \citet{bessell98}.

\begin{figure}
        	\includegraphics[width=\textwidth, angle=-90,trim={1cm 0.5cm 0.cm 1.2cm},clip, width=\columnwidth]{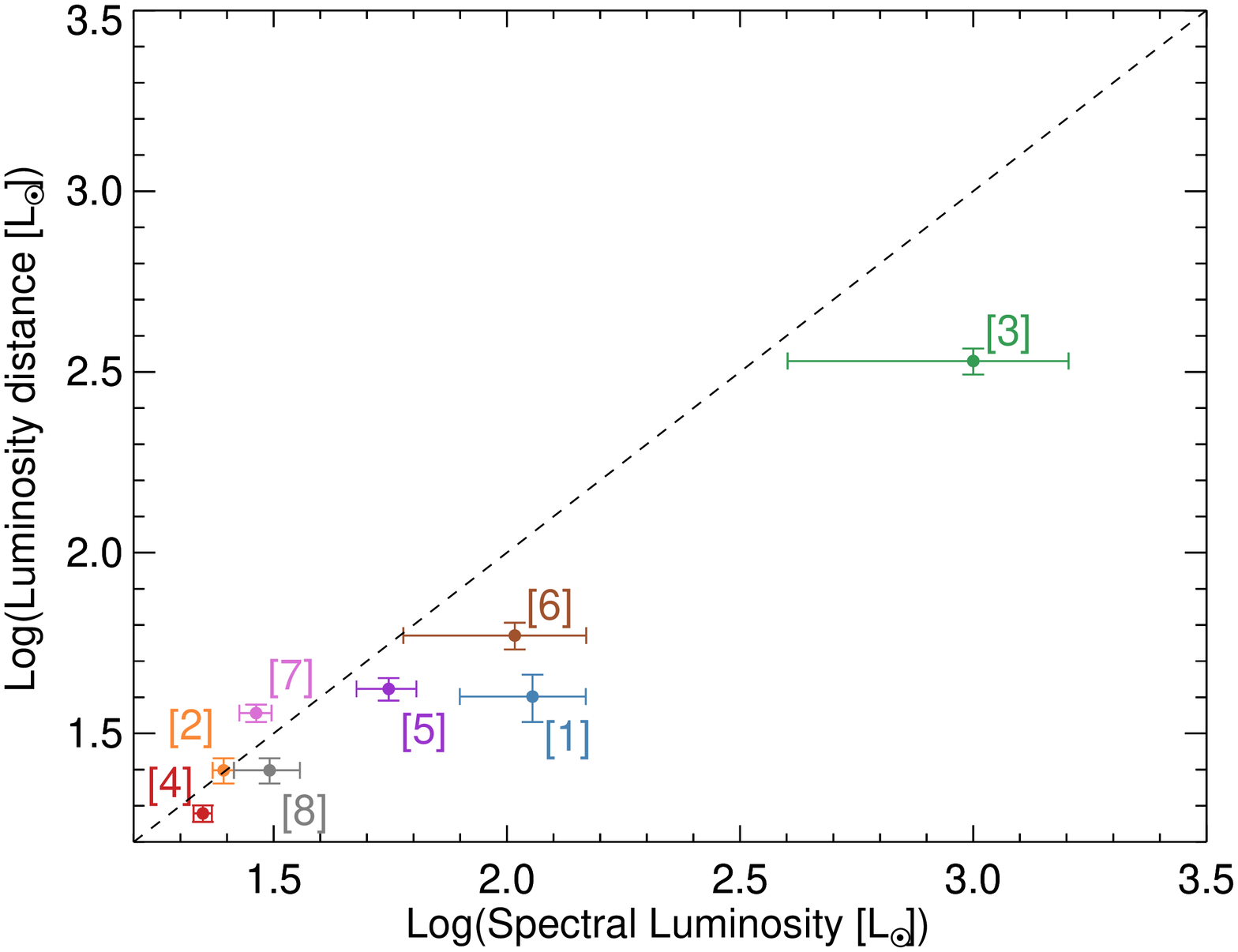}
        	\caption{ Luminosity distance derived from Gaia distance compared with the luminosity found with our spectroscopy analysis. The dashed line is the identity line.}
            \label{fig:luminosity}
\end{figure}

 V magnitudes were used for the calculation and the extinction from the Gaia catalogue in the G filter was transformed to V magnitude using the photometric relationship provided at documentation release 1.1 for Gaia Release 2. Stars [1] and [5] do not have extinction available at Gaia database and this is the main cause of the differences found in luminosity for these stars. The final error bars for the stellar parameters were calculated taking into account the deviations from both parallax and bolometric correction. The results are found in table~\ref{tab:SME_Media_Robusta_vsini_cesam}. Figure~\ref{fig:luminosity} shows the luminosity distance for the main sequence stars derived from Gaia distance compared with the luminosity found with the spectroscopy analysis.
 
\section{Spectroscopic Analysis} \label{spectroscopy}

In order to derive the physical parameters of the stars, spectra were obtained at the Pico dos Dias\,/\,Laborat\'orio Nacional de Astrof\'isica (OPD/LNA, Brazil) with the Cassegrain spectrograph mounted to the Perkin-Elmer Telescope (1.6m), with a first order 1200 lines/mm grating, resulting in a dispersion of $\sim0.5$ \AA/pixel (Fig.~\ref{fig:spectrablue}). During the night of August 26th, 2016, data were obtained in the blue region 4000--5000 \AA with a resolving power of about 9,600 and a signal-to-noise ratio \citep[SNR, computed with][]{stoehr2008} of 270 in average (table~\ref{tab:snr_and_vsini}). In the night of August 28th, observations were made in the red region centered on the H$\alpha$ line,  6562.8 \AA, with a resolving power of about 16,000 and SNR $\sim$ 190. A second stellar spectrum of star [3] was later obtained (July 16th, 2019), with the same instrumental conditions, in the region 3800--4670 \AA \, to check the calcium K line, since our preliminary results conflicted with the Gaia Luminosity determination (Figure~\ref{fig:star3spectra}).
Spectral characterization was made using Spectroscopy Made Easy \citep[SME,][]{Valenti1996}, with initial parameters calculated from those available in the literature to create synthetic spectra. The list of input lines was provided by the Vienna Atomic Line Database \citep[VALD,][]{Heiter2008}. Using interpolated atmospheric models, global stellar parameters were let to vary freely until the best possible fit was achieved (minimization of $\chi^2$).
  
        \begin{figure}
        	\includegraphics[angle=270,width=0.9\columnwidth]{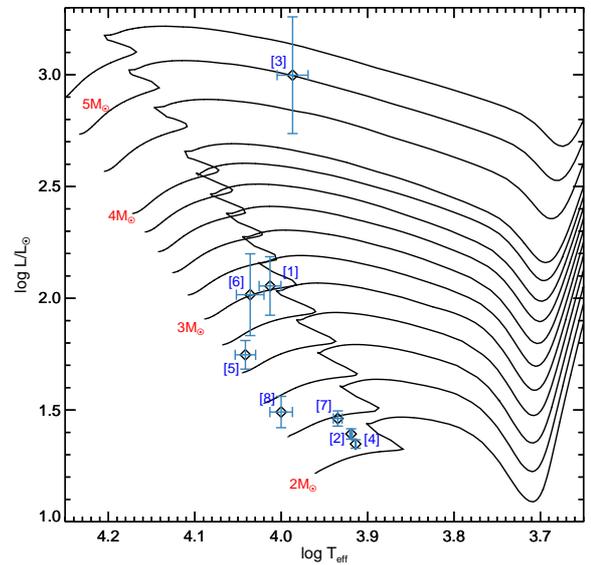}
        	\caption{Evolutionary tracks for A-type stars using the CESAM+POSC grids \citep{Marques2008}. Crosses represent the uncertainties in the parameters derived using the SME code.}
            \label{fig:hrdiagram}
        \end{figure}

\begin{table*}
	\caption{Atmospheric and structural parameters for the stars in table~\ref{tab:target_list}, derived from spectroscopic data and applying SME and the CESAM+POSC grids. The last two columns contains the distance and luminosity derived from Gaia \citep{bailer2018}.}
	\label{tab:SME_Media_Robusta_vsini_cesam}
	\begin{tabular}{lcccccccc} 
		\hline
		Target          & Spectral & $T_\mathrm{eff}$ & $\log g$ & $M$/M$_{\sun}$ & $L$/L$_{\sun}$ & $R$/R$_{\sun}$  & distance & $L$/L$_{\sun}$ \\
		                & Type & [K] & [dex] & & & & [pc] (Gaia) & (Gaia) \\
		\hline
	    $[1]$ & B9V & $10300\pm300$ & $3.9\pm0.1$   & $3.1\pm0.2$ & $114\pm34$ & $3.3\pm0.4$ & $595\pm39$ & $40\pm6$\\
        $[2]$ & A6V & $8300\pm30$   & $4.00\pm0.01$ & $2.10\pm0.03$ & $24.7\pm1.3$   & $2.4\pm0.1$   & $458\pm17$ & $25\pm2$\\
        $[3]$ & A0III/IV & $9700\pm400$  & $3.0\pm0.2$   & $4.9\pm0.9$   & $1000\pm600$  & $12\pm4$  & $559\pm22$ & $339\pm28$\\
        $[4]$ & A7V & $8200\pm10$   & $4.01\pm0.02$ & $2.05\pm0.02$ & $22.3\pm1.0$   & $2.36\pm0.07$ & $306\pm9$  & $19\pm1$\\
        $[5]$ & B9V & $11000\pm300$ & $4.25\pm0.04$ & $2.6\pm0.1$   & $55.8\pm8.2$   & $2.0\pm0.1$   & $469\pm14$ & $42\pm3$\\
        $[6]$ & A0V & $10900\pm400$  & $4.0\pm0.2$   & $3.0\pm0.3$   & $104\pm44$    & $2.6\pm0.6$   & $216\pm4$  & $59\pm5$\\
        $[7]$ & A5V & $8600\pm100$  & $4.01\pm0.02$ & $2.2\pm0.1$   & $29.0\pm2.3$   & $2.4\pm0.1$   & $379\pm10$ & $36\pm2$\\
        $[8]$ & B9V & $10000\pm300$ & $4.5\pm0.1$   & $2.3\pm0.1$   & $31\pm5$       & $1.75\pm0.03$ & $423\pm12$ & $25\pm2$\\
		\hline
	\end{tabular}
\end{table*}

SME output yields the following stellar parameters: effective temperature ($T_\mathrm{eff}$), surface gravity ($\log g$), metallicities, macro and microturbulence velocity, radial velocity, and $v\sin{\textrm{i}}$ and a set of uncertainties. SME provides uncertainty values that are purely numerical and therefore do not properly represent the real uncertainties. Thus, for each adjusted parameter a hundred Monte Carlo simulations were executed \citep{MylesMonteCarlo1996}, randomly varying by 5\% the best fit obtained with SME \citep{Niemczura14} and comparing the results. Some of the uncertainties found have small values. As the SNR and resolving power of the individual spectra are about the same, the Monte Carlo simulation failed to return a realistic value for the deviations. In some cases, the algorithm found solutions only around a local minimum at the parameter space. Values for the uncertainties in stellar parameters deduced from our observations are around 300 K for temperature and 0.1 dex for $\log g$. We assumed as uncertainty in the measure the dispersion 1$\sigma$ of the value of each parameter.

By looking for the values of $T_\mathrm{eff}$ and $\log g$ from SME in the output of evolutionary tracks computed from the Code d'\'Evolution Stellaire Adaptatif et Modulaire (CESAM) and Porto Oscillations Code (POSC) grids \citep{Marques2008}, we were able to find the best structure models -- including mass, radius, and luminosity -- for the eight stars (see table~\ref{tab:SME_Media_Robusta_vsini_cesam} and figure~\ref{fig:hrdiagram}).

\subsection{Rotation} \label{rotation}

\begin{table*} 
	\caption{Signal-to-noise ratio for observed spectra of individual targets, and rotational velocities derived using the two methods described in section~\ref{rotation}. FWHM values were corrected for the instrumental width ($\sim31$ km s$^{-1}$).}
	\label{tab:snr_and_vsini}
	\begin{tabular}{lllcccc} 
		\hline
$\#$ & SNR & SNR & $v\sin{\textrm{i}}$\,(km s$^{-1}$) & $v\sin{\textrm{i}}$\,(km s$^{-1}$) & $\sigma$\,(km s$^{-1}$) &   $v\sin{\textrm{i}}$\,(km s$^{-1}$)\\
       & [blue] & [red] & [Fourier] & [FWHM] & [FWHM] &  SME\\ 
\hline
$[1]$ & $263$   &   $185$  & $127$ & $161$ & $20$ & $169$\\
$[2]$ & $225$   &   $174$  & $141$ & $147$ & $13$ & $163$\\
$[3]$ & $311$   &   $265$  & $143$ & $132$ & $28$ & $167$\\
$[4]$ & $311$   &   $227$  & $143$ & $143$ & $14$ & $150$\\
$[5]$ & $307$   &   $253$  & $161$ & $143$ & $34$ & $158$\\
$[6]$ & $475$   &   $332$  & $169$ & $157$ & $22$ & $180$\\
$[7]$ & $279$   &   $230$  & $140$ & $109$ & $25$ & $176$\\
$[8]$ & $275$   &   $194$  & $142$ & $139$ & $16$ & $145$\\
\hline
\end{tabular}
\end{table*}

Dwarf A-type stars are relatively fast rotators, showing  $v\sin{\textrm{i}}$ values significantly greater than  $100$ km s$^{-1}$ \citep{1999A&AS..137..273G}. Therefore, it is possible to estimate the rotation velocities of our sample from spectral line broadening. Besides to the value of $v\sin{\textrm{i}}$ calculated from SME, we checked the rotation using both a Fourier technique \citep{1976PASP...88..809S} and the measure of the full width at half maximum (FWHM) of a Gaussian fit for the \ion{Mg}{ii} 4481 \AA\,line \citep{gray1976}. The results are summarized in table~\ref{tab:snr_and_vsini}.

\section{Results} \label{results}
 
Only stars [1], [2] and [3] show significant periodicity in the light curves. We present their photometric analysis below.
 
\subsection{Star [1]} \label{results:star1}
  We find $T_\mathrm{eff}$ = $10300\pm300$ K and $\log g$ = $3.88\pm0.10$ dex.
  The star is then a B9\,V with $L = 113.5\pm34.2$ L$_{\sun}$, $R$ = $3.3\pm0.4$ R$_{\sun}$ and $M$ = $3.1\pm0.2$ M$_{\sun}$ (Table \ref{tab:SME_Media_Robusta_vsini_cesam}). \cite{Huber2016} classified this star as colder than our spectroscopy observation indicates. Its parallax of $1.65\pm0.11$ mas was determined by the \citet{gaiacollab}. A distance of $595\pm39$ pc was inferred by \cite{bailer2018}.
  
\begin{figure}
        	\includegraphics[angle=0, trim={.5cm 1cm .5cm .5cm},clip, width=\columnwidth]{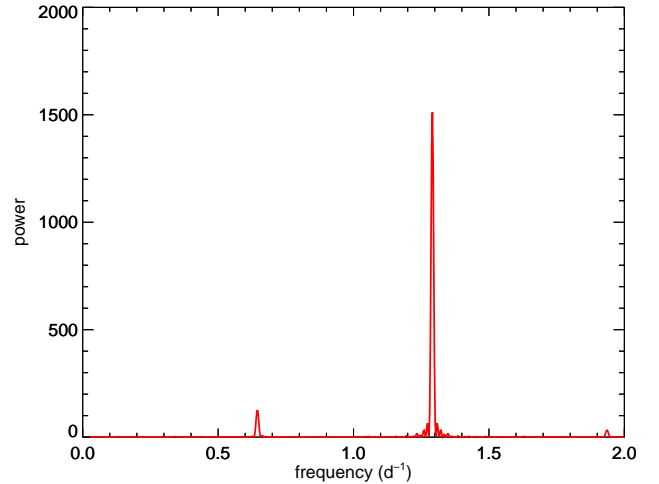}
            \caption{CLEANEST Spectrum of star [1]. The frequency $1.29\pm0.01$ d$^{-1}$ is the first harmonic of $0.64\pm0.01$ d$^{-1}$. The variation is consistent with a rotation of $v\sin{\textrm{i}} = 127\pm22$ km s$^{-1}$ deduced from the spectroscopic observations (see section~\ref{results:star1}).}
            \label{fig:220679442cl}
 \end{figure}

\begin{figure}
        	\includegraphics[width=\linewidth, angle=0,trim={.4cm .5cm .5cm .5cm},clip, width=\columnwidth]{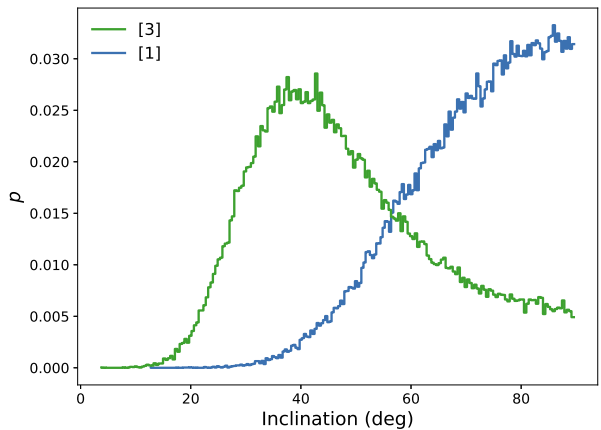}
               \caption{Posterior probability density functions for inclinations of star [1] and [3] calculated with the code \emph{inclinations}.  \citep{White2017}.}
            \label{fig:inclination}
 \end{figure}

  A consistent value of $v\sin{\textrm{i}} = 127\pm22$ km s$^{-1}$ was found by our combined methods.  The CLEANEST spectrum for the photometric observations of this star is shown in Figure~\ref{fig:220679442cl}. The most pronounced frequency is $1.29\pm0.01$ d$^{-1}$, which is the first harmonic of $0.64\pm0.01$ d$^{-1}$ that also appears in the periodogram. The stellar light curve can be well fitted by a sum of sine functions, with $0.64\pm0.01$ d$^{-1}$ and four more harmonics. The fundamental frequency found is consistent with the stellar rotation. 
  \cite{balona2011} interpreted similar B-type stars light curves as due to binarity or rotation. Taken the $v\sin{\textrm{i}}$ and radius values found in this paper and $1\sigma$ uncertainties, the star inclination is estimated to be between $i=60$ and $i=77\degree$. Figure \ref{fig:inclination} shows the posterior probability density function for the inclination \citet{White2017}. 
  The frequency at $1.29 \pm 0.01$ d$^{-1}$ appears as the strongest in the power spectrum. It is usual to find half of the rotation period in stellar light curves. In this case, as the fundamental frequency has a smaller amplitude than the 2nd harmonic, a possible scenario would be the presence of two spots on the stellar surface in anti-phase \citep{collier2009,walkowicz2013}.

\subsection{Star [2]} \label{results:star2}
The spectroscopy analysis indicates $T_\mathrm{eff}$ = $8300\pm30$ K and $\log g$ = $4.00\pm0.01$ dex (Table~\ref{tab:SME_Media_Robusta_vsini_cesam}). The star is a A6\,V with $L = 24.7\pm1.3$ L$_{\sun}$, $R = 2.4\pm0.1$ R$_{\sun}$ and $M = 2.10\pm0.03$ M$_{\sun}$. These values are greater than those found in \cite{Huber2016}. Its parallax of $2.15\pm0.08$ mas was determined by \citet{gaiacollab}. A distance of $458\pm17$ pc was thus inferred by \cite{bailer2018}. Our spectroscopic data indicate $v\sin{\textrm{i}} = 147\pm13$ km s$^{-1}$. Given its mass,  $T_\mathrm{eff}$ and luminosity, this star is at the upper limit of the $\delta$ Scuti instability strip region in the HR diagram.

Its measured $v\sin{\textrm{i}}$ is typical for A and F stars, and close to their observed mean value of 134 km s$^{-1}$ as in \cite{2015MNRAS.450.2764N} and well inside the range estimated by \citet{2012A&A...537A.120Z}. 

\begin{table*}
\footnotesize
\centering
	\caption{Frequencies, amplitudes, and phases obtained using PERIOD04 for star [2]
(columns 2, 4, and 7 respectively). It is also shown the standard deviation of the frequencies (3rd column) and the mean amplitude (5th and 6th columns) over 1000 realizations after adding random noise to our model (see text).}
	\label{tab:freq6722}
	\begin{tabular}{ccccccc}
		\hline
		 Frequency & Frequency & Frequency error & Amplitude & Mean Amplitude & Mean Amplitude & Phase $\phi_i$ \\
		  $\#$ & $\nu_i$ (d$^{-1}$) & (d$^{-1}$) & $A_i$ (ppm) & (ppm) & ($\sigma$) & (radians) \\
		\hline
01 &  16.3569 & 0.0001 &  679. &  671. &  116. &  0.19 \\
02 &  14.3903 & 0.0001 &  676. &  673. &  112. &  0.73 \\
03 &  11.7737 & 0.0001 &  594. &  637. &  102. &  0.04 \\
04 &  13.2736 & 0.0001 &  578. &  584. &   91. &  0.25 \\
05 &  16.8177 & 0.0001 &  523. &  544. &   92. &  0.39 \\
06 &  14.1294 & 0.0001 &  382. &  364. &   63. &  0.17 \\
07 &  15.7061 & 0.0002 &  329. &  300. &   54. &  0.90 \\
08 &  16.9111 & 0.0001 &  300. &  326. &   55. &  0.17 \\
09 &  16.9387 & 0.0002 &  292. &  291. &   50. &  0.28 \\
10 &  10.9609 & 0.0002 &  247. &  255. &   41. &  0.35 \\
11 &  11.7517 & 0.0001 &  208. &  294. &   47. &  0.56 \\
12 &  15.9993 & 0.0002 &  199. &  199. &   32. &  0.42 \\
13 &  18.9606 & 0.0002 &  193. &  184. &   30. &  0.02 \\
14 &  13.2439 & 0.0003 &  187. &  208. &   34. &  0.67 \\
15 &  18.1723 & 0.0003 &  176. &  166. &   28. &  0.71 \\
16 &  16.5065 & 0.0004 &  131. &  129. &   22. &  0.38 \\
17 &   9.3736 & 0.0004 &  112. &  117. &   19. &  0.09 \\
18 &  13.7562 & 0.0004 &  106. &  106. &   17. &  0.14 \\
19 &  20.5399 & 0.0006 &   79. &   77. &   13. &  0.91 \\
20 &  19.3717 & 0.0011 &   79. &   62. &   11. &  0.76 \\
21 &  10.9139 & 0.0008 &   75. &   67. &   11. &  0.20 \\
22 &  20.1277 & 0.0006 &   72. &   75. &   13. &  0.08 \\
23 &  12.4666 & 0.0007 &   64. &   55. &    9. &  0.73 \\
24 &  14.9863 & 0.0008 &   58. &   56. &    9. &  0.62 \\
25 &  12.6566 & 0.0006 &   54. &   58. &    9. &  0.84 \\
26 &  21.0003 & 0.0004 &   51. &   92. &   16. &  0.22 \\
27 &  22.0189 & 0.0010 &   46. &   42. &    8. &  0.01 \\
28 &  12.4755 & 0.0014 &   40. &   36. &    6. &  0.94 \\
29 &  13.4488 & 0.0019 &   40. &   35. &    6. &  0.11 \\
30 &  11.4116 & 0.0017 &   38. &   35. &    6. &  0.90 \\
31 &  12.0024 & 0.0017 &   37. &   40. &    7. &  0.84 \\
32 &  14.0901 & 0.0005 &   36. &   75. &   12. &  0.15 \\
33 &  21.0603 & 0.0011 &   36. &   42. &    7. &  0.12 \\
34 &  18.3205 & 0.0012 &   32. &   41. &    7. &  0.20 \\
35 &  12.0305 & 0.0020 &   29. &   45. &    7. &  0.16 \\
36 &  23.0596 & 0.0016 &   28. &   34. &    6. &  0.94 \\
37 &  17.4680 & 0.0015 &   27. &   33. &    6. &  0.04 \\
38 &  10.3004 & 0.0008 &   27. &   56. &    9. &  0.97 \\
39 &  22.6547 & 0.0011 &   26. &   38. &    7. &  0.06 \\
40 &  14.3518 & 0.0029 &   23. &   89. &   12. &  0.11 \\
		\hline
		\end{tabular}
\end{table*}

\begin{figure*}
        	\includegraphics[angle=0,width=\textwidth]{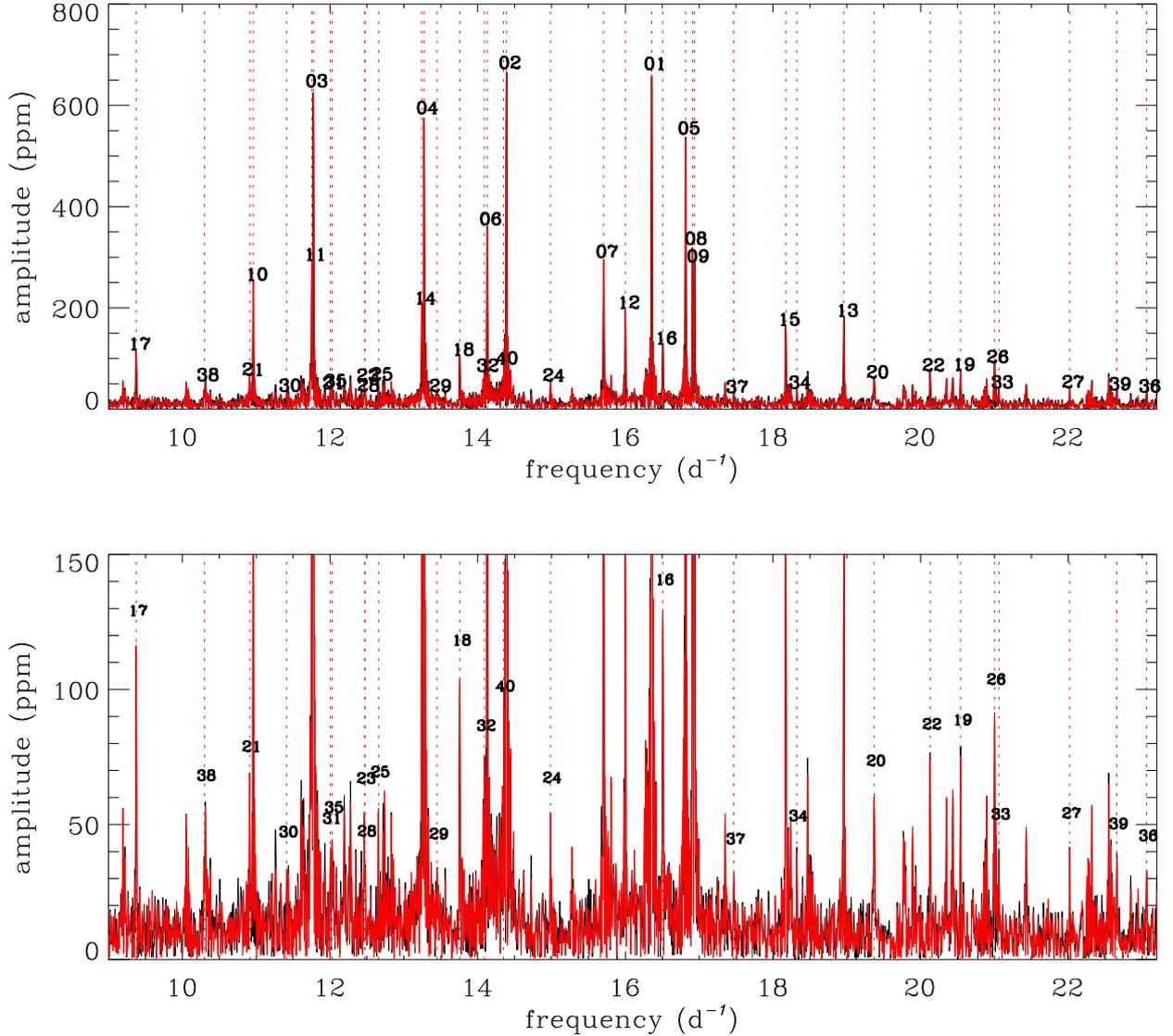}
            \caption{The power spectrum of star [2] calculated with Periodo4 (in black) and the model spectrum over it in red. The observed one is barely visible as the spectrum of the model has a very good resemblance to it. 
            At the top panel we show all observed peaks, while at the bottom only the fitted frequencies with amplitudes smaller than 150 ppm are shown for better visualization. The red vertical dotted lines in both panels correspond to the fitted frequencies and the corespondent numbers in Table~\ref{tab:freq6722} are plotted on top of each peak.}
            \label{fig:6722obsspec}
 \end{figure*}

Figure \ref{fig:6722obsspec} shows the observed power spectrum for the photometric observations of this star (shown in panel 2 of Figure~\ref{fig:allphotometry}).
The variability of star [2] is typical of a $\delta$ Scuti star.
Using PERIOD04 \citep{Lenz2005}, we obtain forty frequencies along with their amplitudes and phases.

The filling factor of {\it{Kepler}} data is 0.9 and the amplitude spectrum was multiplied by 1.118 as in \cite{komm2000} to correct for it. 

We model the photometric variation as: 
\begin{equation}
   \sum_{i=1}^{i=40} A_i \sin(2\pi \nu_i t + \phi_i), 
\end{equation}

where $A_i$, $\nu_i$, and $\phi_i$ are the individual amplitudes, frequencies and phases respectively (see Table~\ref{tab:freq6722}). We use the same observed window ($t$) as our data (including the number, position, and length of the gaps). The spectrum of our model fits very well the observed one and is shown in red in Figure~\ref{fig:6722obsspec}. 

The observed modes have a line width smaller than the frequency resolution of the observed time series, which indicates a long lifetime of the modes.

\begin{figure*}
        	\includegraphics[angle=0,width=\textwidth]{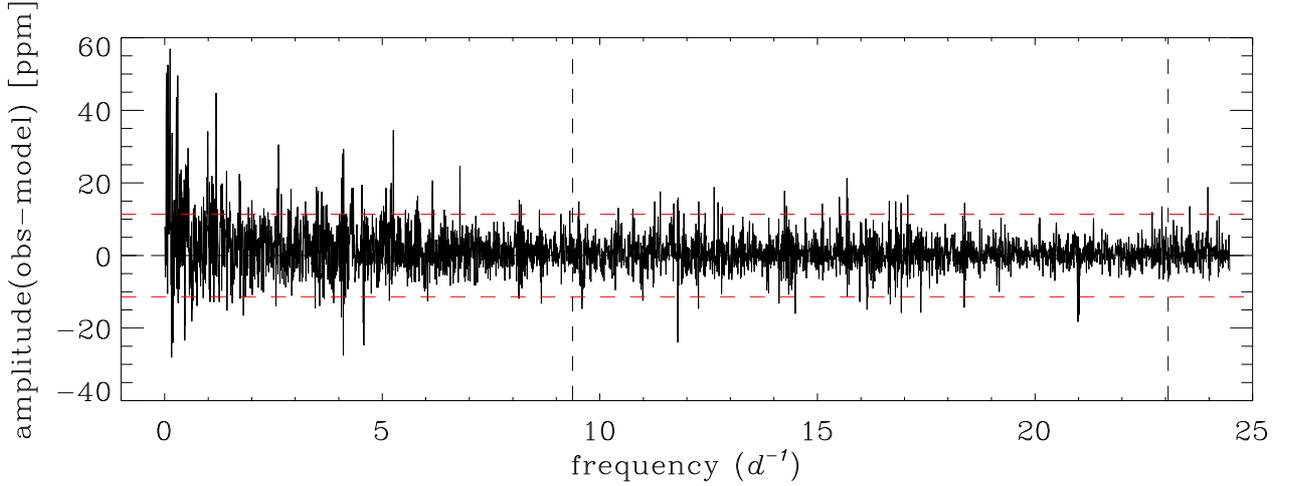} 
            \caption{Residual spectrum of star [2] where the model spectrum was subtracted from the observed spectrum. The horizontal dashed red lines correspond to three times the standard deviation. The vertical dashed lines correspond to the lowest and largest frequency in Table~\ref{tab:freq6722}. The residuals at lower frequencies do not indicate the presence of $g$ modes.
            }
            \label{fig:6722diffspec}
 \end{figure*}

To check for the presence of the 40 frequencies in the light curve, we performed an F-test. This statistical test calculates the ratio if the sum of residuals squared decrease significant more than the relative change of the degrees of freedom from a simple model to a more complex one. This test shows that all 40 frequencies are statistically significant at 99$\%$ confidence level.
The noise calculated as the average amplitude at frequencies higher than 5 d$^{-1}$ after prewhitening all 40 frequencies is equal to 5 ppm. \citet{breger1993} chose a signal-to-noise ratio of 4 as a good criterion to distinguish between peaks due to pulsation and noise, which applies to our results in Table~\ref{tab:freq6722} where all amplitudes are larger than 20 ppm.
Figure~\ref{fig:6722diffspec} shows the difference between the observed spectrum and that of our model (black and red lines in Figure~\ref{fig:6722obsspec}). The standard deviation of the residuals for frequencies higher than 5 d$^{-1}$ is equal to 4.0 ppm and there is no obvious peak present.
Figure~\ref{fig:6722diffspec} shows the residuals for lower frequencies as well. This $\delta$ Scuti-type star seems to have only $p$ modes with no hybrid $p$ and $g$ modes since no significant frequencies below 5 d$^{-1}$ are seen \citep{bowman_kurtz2018}.

We added random noise to our model with a standard deviation given by the time-series residuals calculated by PERIOD04, which is equal to 239 ppm. 

The amplitude spectrum with an oversampling factor of 16 was calculated for 1000 realizations. 

For each realization, we obtained the forty frequencies and their amplitudes by fitting a Gaussian function at each peak. We then calculated the average and standard deviation of the amplitude and frequency of each mode over 1000 realizations. 
The standard deviation of each peak amplitude varies from 5.4 to 6.3 ppm. The mean amplitude of each mode amplitude in ppm and divided by its standard deviation, $\sigma$, is given in Table~\ref{tab:freq6722} (5th and 6th columns, respectively). Seventy percent of the observed amplitudes are within 3$\sigma$ of its mean amplitude.
The standard deviation of each frequency is also given in 
Table~\ref{tab:freq6722} (3rd column) and gives an estimation of the frequency uncertainty. The frequency resolution of the spectra with a 16-factor oversample is 0.0008 d$^{-1}$. 
The mean frequency of each peak is very close to the observed one and it is not shown.

\begin{figure}
    \centering\includegraphics[width=\columnwidth]{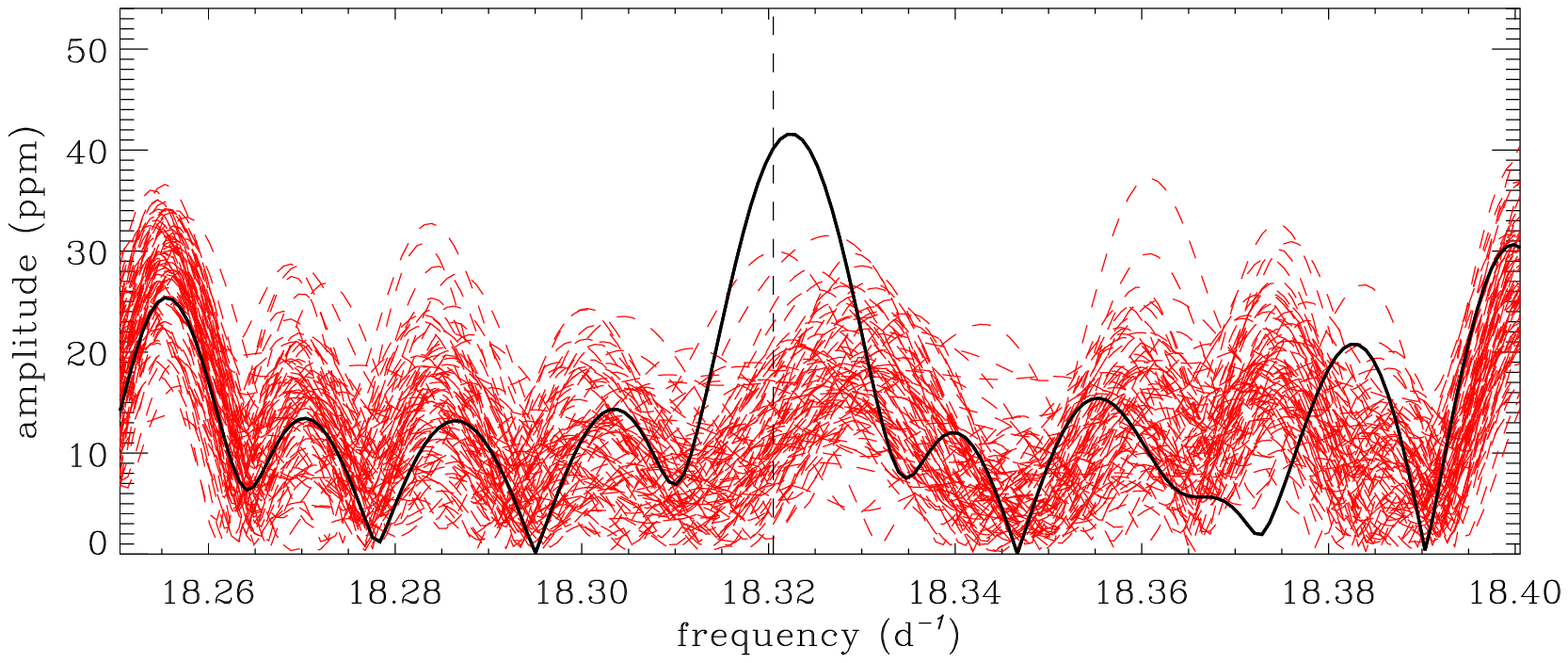}
    \centering\includegraphics[width=\columnwidth]{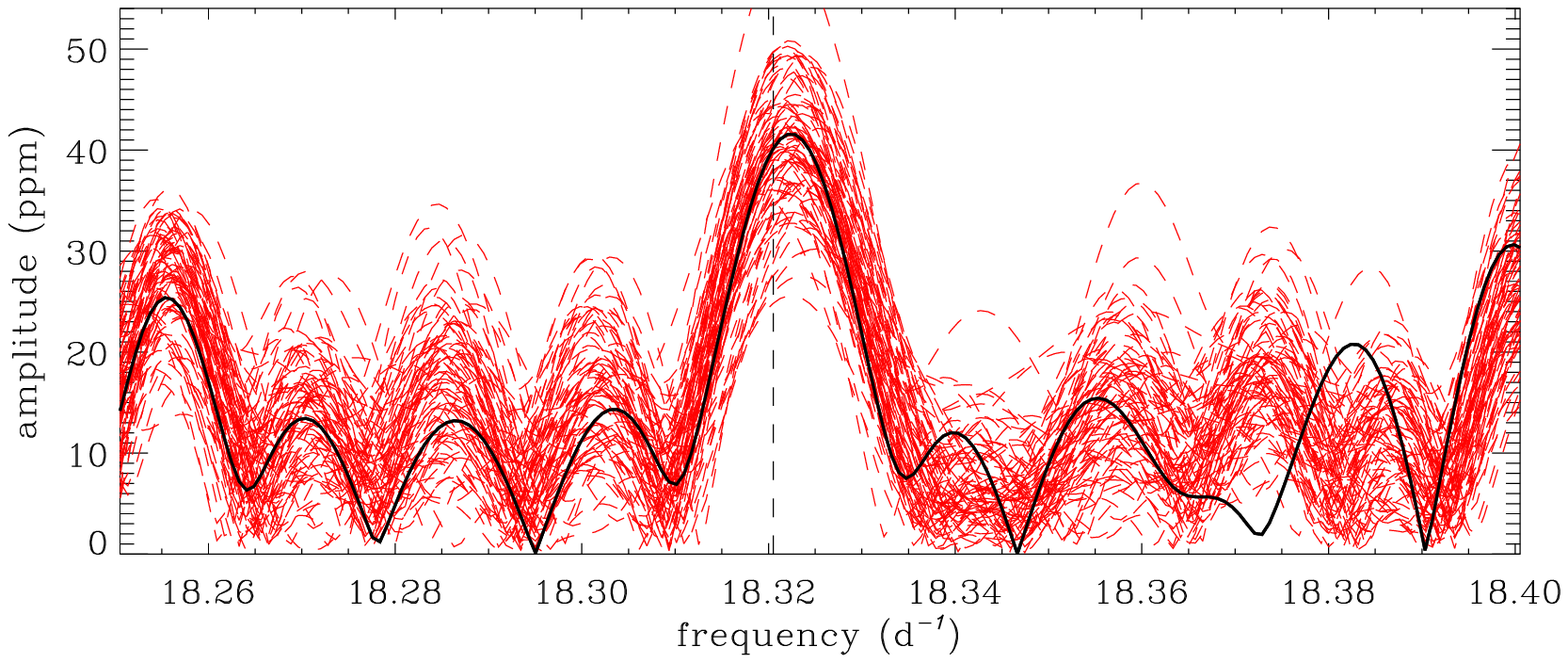}
  \caption{Example of fitting one of the peaks with small amplitude in Table~\ref{tab:freq6722}. The black line is the observed spectrum in both panels. The red lines correspond to 100 realizations without including peak number 34 (top panel) and including it (bottom panel).}
  \label{fig:6722peak}
 \end{figure}

Figure~\ref{fig:6722peak} shows an example of the significance of fitting the peaks with small amplitudes.

Modes number 1, 12, 16, 19, 22, 23, 28, 31 and 35 in Table~\ref{tab:freq6722} have frequencies close to multiple integers of 4.08 d$^{-1}$, i.e., to harmonics of the frequency
introduced in the data by the {\it{Kepler}} satellite thruster firings that occur about every six hours. Despite the correction used in the light curve, there still might be some small effect.
Our minimum and maximum observed frequencies, 9.3736 d$^{-1}$ and 23.0596 d$^{-1}$ (i.e., 108.491 and 266.894 $\mu$Hz) agree with the theoretical values obtained by \citet{Michel2017} for a 2.0 solar mass star, as illustrated in their Figure 2.
The amplitudes of the oscillations are in the expected range as obtained by \citet{Michel2017} using the CoRoT data and defining the square root of the quadratic sum of amplitudes of all peaks as the amplitude index. For our results, the amplitude index is equal to 1600 ppm.

 We can not rule out that some of the frequencies in Table~\ref{tab:freq6722} are combination frequencies
 due to non-linearity of pulsation modes \cite[see][and references therein]{kurtz2015}. Although the observed frequencies are in the $p$ mode frequency range and their amplitudes are lower than 700 ppm, they are close to some cases described in \cite{kurtz2015}.
 As stellar rotation lifts the degeneracy of non-radial pulsation modes into its 2$\ell$+1 components,
we would expect the rotational splittings to be equal or larger than 1.21 d$^{-1}$ given our determinations of the $v\sin{\textrm{i}}$ and stellar radius.
For moderate and fast rotating stars, 
the splitting will be significantly asymmetric due to the Coriolis force \citep{2010aste.book.....A}, making their identification difficult.

To search for frequency spacings, we performed a Fourier analysis of our frequencies, assuming unit amplitude for all peaks, but the observed peaks are not statistically significant as to give coherent results.

Although we were unable to identify rotational splittings or the large separation,
17 out of the 40 identified frequencies are equispaced by 1.23 d$^{-1}$ or a multiple of it.

The number of separations between the identified frequencies for one, two, three and four times
1.23 d$^{-1}$ are, respectively, four, five, five and three.

The average and its standard deviation of all 17 separations divided by its multiplication factor is equal to 1.230$\pm$0.005 d$^{-1}$.
\citet{suarez2014} estimated a linear relation between the large separation and the mean density of $\delta$ Scuti stars from asteroseismic models. Given the values in Table~\ref{tab:SME_Media_Robusta_vsini_cesam}, star [2] has a mean density $\bar{\rho}$ = 0.15 $\pm$ 0.02 $\bar{\rho}_{\sun}$.
The large separation predicted by \citet{suarez2014} is approximately inside the 3.4-4.0 d$^{-1}$ interval, which agrees quite well with three times 1.23 d$^{-1}$.

 \subsection{Star [3]} \label{results:star3}
 
\begin{figure}
        	\includegraphics[angle=90,trim={.5cm 1.5cm 1cm 1cm},clip, width=\columnwidth]{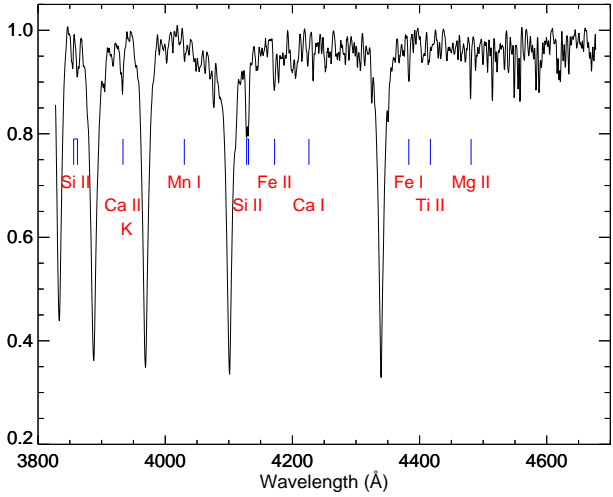}
            \caption{Second spectrum of star [3] observed with the 1.6m P-E telescope in OPD/LNA on the night of July 16th, 2019, featuring the spectral classification region (3800--4670\AA). The main lines used for physical parameters estimation are identified.}
            \label{fig:star3spectra}
 \end{figure}

  In our spectroscopy analysis we found $T_\mathrm{eff}$ = $9700\pm400$ K and $\log g$ = $3.0\pm0.2$  dex (Table \ref{tab:SME_Media_Robusta_vsini_cesam}).
  The star is a A0\,III/IV with $R$ = $12\pm4$ R$_{\sun}$, $M$ = $4.9\pm0.9$ M$_{\sun}$ and $L$ = $1000\pm600$ L$_{\sun}$. Its parallax of $1.76\pm0.07$ mas was determined by \citet{gaiacollab}. The distance of $559\pm22$ pc was inferred by \cite{bailer2018}. \citet{Catanzaro19} found $T_\mathrm{eff}$ = $8000\pm125$ K and $\log g=3.50\pm0.25$ dex and \citet{Huber2016} found $T_\mathrm{eff}$ = $8803\pm125$ K and $\log g=3.73\pm0.20$ dex. Therefore since the Ca K line is weak (Figure~\ref{fig:star3spectra}), this indicate star [3] is hotter than the temperature found by \citet{Catanzaro19} and \citet{Huber2016}. The value of $v\sin{\textrm{i}}=143 \pm 28$ km s$^{-1}$  is consistent with all our determinations displayed at table~\ref{tab:snr_and_vsini}.
  
\begin{figure}
        	\includegraphics[angle=0,trim={.5cm 1cm .5cm .5cm},clip, width=\columnwidth]{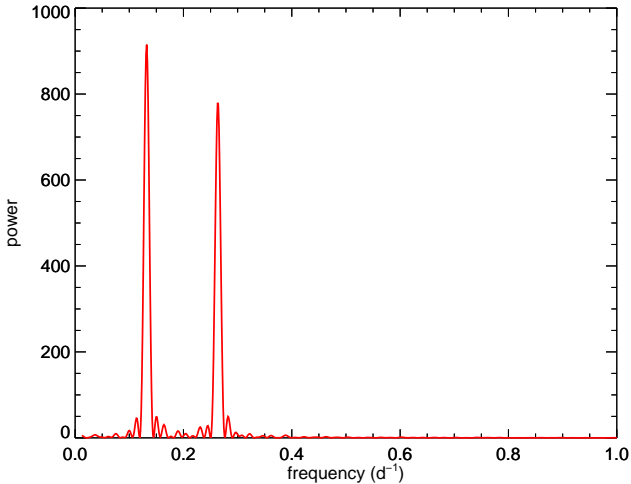}
            \caption{CLEANEST Spectrum of star [3]. The most pronounced frequencies are $0.13\pm0.01$ d$^{-1}$ (fundamental) and the first harmonic of $0.26\pm0.01$ d$^{-1}$.  The frequency found is consistent with rotation.}
            \label{fig:220532854cl}
 \end{figure}

\begin{figure}
  
        	\includegraphics[angle=90,trim={.1cm .8cm .8cm 3cm},clip,width=\columnwidth]{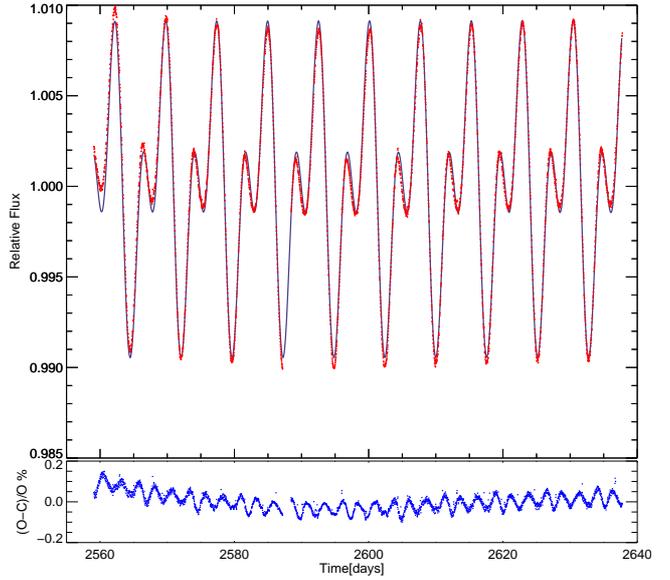}
            \caption{Top: Fit of star [3] light curve with the two main frequencies (solid line). Bottom: Residuals (O-C) after removing the above fitted function.}
            \label{fig:220532854fit}
 \end{figure}

\begin{figure}
            \includegraphics[width=\linewidth, angle=-90,trim={.8cm 2cm 1.5cm .5cm},clip, width=\columnwidth] {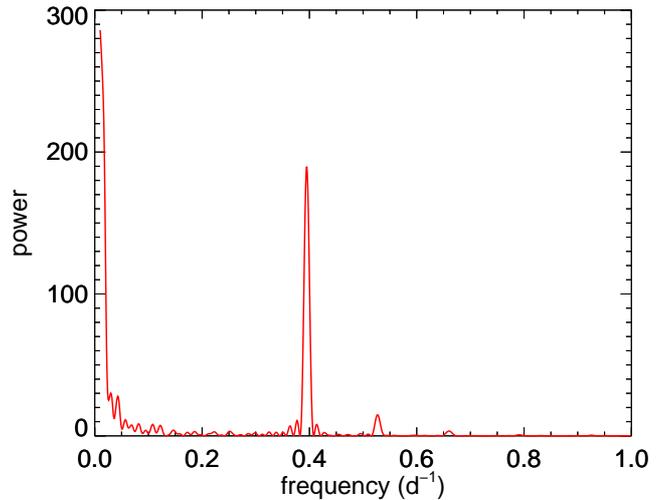}
            \caption{CLEANEST spectrum of star [3], after removing the two main frequencies.}
            \label{fig:220532854rcl}
 \end{figure}

  Figure~\ref{fig:220532854cl} shows the CLEANEST spectrum for the photometric observations of this star. The most pronounced frequencies are $0.13\pm0.01$ d$^{-1}$ (fundamental) and the first harmonic of $0.26\pm0.01$ d$^{-1}$. 
 The frequency found at CLEANEST Spectrum is consistent with rotation. Taken the values of $v\sin{\textrm{i}}$ and radius found in the paper and $1\sigma$ uncertainties,  the star inclination is between $i=27\degree$ and $i=66\degree$. Figure \ref{fig:inclination} shows the posterior probability density function for the inclination \citep{White2017}. A model with the two frequencies mentioned above give an excellent fit to the light curve.    
 (Figure~\ref{fig:220532854fit}). However, the CLEANEST spectrum of the residual (O-C) reveals the third harmonic at $0.39\pm0.01$ d$^{-1}$ and a long term variation (Figure~\ref{fig:220532854rcl}). \citet{McNamara2012} classified similar light curves as binary / rotation modulated stars (See figure 8 in that paper).  
 
 Spectroscopy data reveals a well-defined silicon doublet (\ion{Si}{ii} 4128-4131 \AA) and (\ion{Si}{ii} 3856-3862 \AA), a characteristic presented by the peculiar magnetic Ap stars and already reported by \cite{2009A&A...498..961R} in their catalog of peculiar A stars. However, in our spectral analysis, we find that star [3] is an evolved object -- a luminous giant star --, which does not correspond to an Ap class star, usually found in the Main Sequence branch \citep{North...1997, 1997A&A...wade}. In figure~\ref{fig:spectrablue} it can be seen that star [3] shows hydrogen lines narrower than the other stars in the sample, suggesting a giant classification.
 
\section{Discussion and Summary} \label{summary}

This paper presents a thorough analysis of massive stars observed photometrically near $50\degree$ below the galactic plane by the {\it{Kepler}}/K2 mission and spectroscopically at OPD/LNA (Brazil). 

We estimated the distance of our stars from the galactic plane. It ranges from 326 pc for star [8] up to 469 pc for star [1]. Since the Milky Way's thin disk has a vertical height scale of 300$\pm$50 pc in the vicinity of the Sun, these stars are at the edge of the old thin disk or inside the thick disk. B stars are confined to the thin young disk extending only about 50 pc below and above the plane.

Three of our targets show significant periodicity. Stars [1] and [3] show periods linked to the stellar rotation. Star [2] is a $\delta$ Scuti variable in which 40 individual frequencies were found by our analysis. Star [3] has a pronounced silicon doublet (\ion{Si}{ii} 4128-4131 \AA), characteristic of the peculiar magnetic Ap stars. However, our spectral analysis showed that it is an evolved star, which does not agree with the Ap class characteristics. This object is in fact a luminous giant, while Ap stars are found in the Main Sequence. Stellar luminosities were calculated both using CESAM+POSC grids and Gaia distance data. For the main sequence stars the results are in agreement at a 2$\sigma$ level.  For stars [4] to [8], no significant periodicity was found in the light curves. Their physical parameters are presented in Table~\ref{tab:SME_Media_Robusta_vsini_cesam} and their rotational velocities estimates for both Fourier and FWHM methods are shown in Table~\ref{tab:snr_and_vsini}.

\section*{Acknowledgements}

This study was financed in part by the Coordena\c{c}\~ao de Aperfei\c{c}oamento de Pessoal de N\'ivel Superior - Brazil (CAPES) - Finance Code 001. This work was also supported with resources from Conselho Nacional de Pesquisa (CNPq) 308871/2016-2 and the State of Paran\'a Secretary of Science, Technology and Higher Education - Fundo Paran\'a. This research was also supported in part by Minas Gerais State Agency for Research and Development (FAPEMIG), Brazil. We also thank S\~ao Paulo Research Foundation (FAPESP) support through grant 2016/13750-6.

This paper includes data collected by the {\it{Kepler}} mission. Funding for the {\it{Kepler}} mission is provided by the NASA Science Mission Directorate.
Some of the data presented in this paper were obtained from the Mikulski Archive for Space Telescopes (MAST). STScI is operated by the Association of Universities for Research in Astronomy, Inc., under NASA contract NAS5-26555. Support for MAST for non-HST data is provided by the NASA Office of Space Science via grant NNX13AC07G and by other grants and contracts.

This paper is also based on observations obtained at the Pico dos Dias Observatory (LNA/MCTIC).

This work has made use of the VALD database, operated at Uppsala University, the Institute of Astronomy RAS in Moscow, and the University of Vienna.

This research has made use of the SIMBAD database, operated at CDS, Strasbourg, France and of data from the European Space Agency (ESA) mission {\it Gaia} (\url{https://www.cosmos.esa.int/gaia}), processed by the {\it Gaia} Data Processing and Analysis Consortium (DPAC, \url{https://www.cosmos.esa.int/web/gaia/dpac/consortium}). Funding for the DPAC has been provided by national institutions, in particular the institutions participating in the {\it Gaia} Multilateral Agreement.




\bibliographystyle{mnras}
\bibliography{refs} 








\bsp	
\label{lastpage}
\end{document}